\documentclass[12pt]{article}
\usepackage[pctex32]{graphics}
\begin{document}
\vspace{3cm}
\begin{center}
{\bf  \Large  On Tachyon Condensation of Intersecting Noncommutative Branes in M(atrix) Theory}
\vspace{1cm}

                      Wung-Hong Huang\\
                       Department of Physics\\
                       National Cheng Kung University\\
                       Tainan,70101,Taiwan\\

\end{center}
\vspace{1cm}
\begin{center}  {\bf  \Large ABSTRACT}\end{center}

The interaction between the intersecting noncommutative D-branes (or membranes) is investigated within the M(atrix) theory.  We first evaluate the spectrum of the off-diagonal fluctuation and see that there is a tachyon mode, which signals the instability of the intersecting branes.  We next explain in detail how the tachyon modes (which are the off-diagonal fluctuations) can be combined with the original branes (which are the diagonal elements) to become a new diagonal matrix, which then represents the new recombined configurations after the tachyon condensation.  The tachyon condensation configurations of the noncommutative branes we find are different from those of the commutative cases studied in the previous literatures.

\vspace{2cm}
\begin{flushleft}
E-mail:  whhwung@mail.ncku.edu.tw\\
\end{flushleft}


\newpage
\section{Introduction}
It is well-known that a tachyon develops when a D-brane and an anti-D-brane come up to some critical interval of order of string scale [1] or a D-brane is intersecting with another D-brane [2,3].  The tachyon fields in these systems will roll down to a stable point.   Sen had argued that the D-brane-anti-D-brane pair is annihilated via the tachyon condensation.  In this process the original D-brane disappears and the kink-type tachyon condensed states corresponding to lower-dimensional D-branes may be produced.   

    D-branes at angles or intersecting D-branes scenario contains several phenomenologically appealing general features.    It provides the construction of explicit D-brane models which at low energy may yield the chiral fermion and gauge group of the standard model [4,5].  It can also provide a simple mechanism for the inflation in the early universe [6].   

    Recently the recombination processes of intersecting D-branes has been investigated in different ways.   The first one [7,8] uses the super Yang-Mills theories which are low energy effective theories of D-branes.   The Yang-Mills field therein represents the dynamics field on the branes while the Higgs fields represent the locations of the D-branes.  The second approach [9] uses the effective tachyon field theory [10], in which the dynamical fields on the branes are the fluctuating in the background of kink solution and the locations of the D-branes are at the zeros of the tachyon field.   The third approach uses the world-volume field theory [12], which can avoid the small angle approximation adopted in the other approaches.  

   In this paper we will study the recombination processes of intersecting D2-branes in the M(atrix) theory [12,13].  It is known that the original M(atrix) theory of BFSS Lagrangian [12] is used to describe the dynamics of D0-branes.  However, as the D-branes can be formed from the M(atrix) theory [14,15] the BFSS Lagrangian can also be used to describe the M-branes and D-branes.  The fact that D0-brane can be described as the magnetic flux of the gauge field living on  D2-branes [15] enables us to describe the dynamics of  D2-branes in the M(atrix) theory.   Since the spacetime coordinates on a D-brane in the presence of non-zero gauge field are actually non-commutative [16],  the intersecting D2-branes described in M(atrix) theory is therefore noncommutative. 

   In section II we follow the method of  [17] to setup the Lagrangian of  intersecting D2-branes in M(atrix) theory.  We adopt the method of [18] to find the spectrum of the off-diagonal fluctuations which represent the interaction between the pair of intersecting noncommutative D-branes (or membranes).   The tachyon modes therein signals the instability of the intersecting branes.  In section III we discuss the effects of the tachyon on the intersecting D2-branes and detail how to find the tachyon condensation configurations of the intersecting noncommutative D-branes.   We see that the recombined configurations are different from those of the commutative cases studied in the previous literatures [7-9].   Final section is devoted to conclusion.

\section {Branes Dynamics: Off-diagonal Fluctuation}
The matrix model we will be concerned with is described by Hamiltonian [12]

$$  H = R ~Tr \left\{{{\Pi_i \Pi_i}\over 2} + {1\over 4}[X_i,X_j]^2
+ \theta^T \gamma_i [\theta,X_i]\right\}, ~~~~ i=1,...,9. \eqno{(2.1)} $$
\\
in which $R$ is the tension of D0 brane.  The configurations of a pair of  D2-branes (or membranes) intersecting  at one angle $\theta$ may be expressed as [17]

$$X_1^0 = \left( \begin{array}{cc}P_1 sin\theta&0\\ 0&P_2 sin\theta  \end{array} \right), ~~~ X_2^0 = \left( \begin{array}{cc}P_1 cos\theta&0\\
0& - P_2 cos\theta  \end{array} \right), ~~~X_3^0 = \left( \begin{array}{cc}
Q_1&0\\ 0&Q_2 \end{array} \right), \eqno{(2.2)}$$
in which 
$$ [Q_1,P_1] = [Q_1,P_1] = 2\pi i z^2, \eqno{(2.3)}$$
where $z^2$ is proportional to the gauge field living on the D2 branes [16] or the charge density of the membrane [14].  The other matrix $X_\mu^0 = 0$, $\mu=4,...,9$.
\\

   When $\theta =0$ and let $X_1^0 = \left( \begin{array}{cc}r&0\\ 0&-r \end{array} \right) $, then we have a membrane and an anti-membrane with distance $2r$ between them.   This system was investigated by Aharony and Berkooz [18].   They computed the long range force therein and found that it is consistent with the supergravity theory.   At short distances the system exhibits the tachyon instability as discussed in [18] and [19].   In this paper we will extend the investigation to the case of general values of $\theta$, and, expecially see how the tachyon condensation will proceed during the recombination of the intersecting branes systems in the matrix theory approach.   Note that the interaction in the angled branes system has also been computed in [17] and [20] in which, however, the branes system they considered do not intersect with each other and therefore has no tachyon instability.

   To proceed we add the following matrix 

$$A_i = \left( \begin{array}{cc} 0&T_i\\ T_i^\dagger &0 \end{array} \right), \eqno{(2.4)}$$
\\
to $X_i^0$, which represents the interaction between the intersecting branes.   In this formulation the fields $T_i$ will be functions of the coordinate $P_j$ and intersection angle $\theta$.  They do not depend on the coordinates $Q_j$ [18].  

   Now, for the cases of $i, j = 1,2,3$, we use the relations

$$ Tr [(X_i^0+A_i), (X_j^0+A_j)] ^2 =  Tr [X_i^0, X_j^0] [X_i^0, X_j^0] +
4 Tr [X_i^0, X_j^0] [X_i^0, A_j] + 2 Tr [X_i^0, X_j^0] [A_i, A_j] + $$ 
$$2Tr [X_i^0, A_j] \left([X_i^0, A_j] + [A_i, X_j^0]\right) + 4 Tr [X_i^0, A_j] [A_i, A_j] + Tr [A_i, A_j] [A_i, A_j],  \eqno{(2.5)}$$
\\
the property of  (2.3) and the method in [18] to find that 

  $$ Tr [X_i^0, X_j^0] [X_i^0, X_j^0] = -16 \pi z^4,  \eqno{(2.6)}$$
  $$ Tr [X_i^0, X_j^0] [X_i^0, A_j] = 0,  \eqno{(2.7)}$$
  $$ Tr [X_i^0, A_j] [A_i, A_j] = 0,  \eqno{(2.8)}$$
 $$Tr [A_i, A_j] [A_i, A_j] = 4\left[(T_1 T_2^\dagger - T_2 T_1^\dagger)^2 + (T_1 T_3^\dagger - T_3 T_1^\dagger)^2 + (T_2 T_3^\dagger - T_3 T_
2^\dagger)^2 \right].\eqno{(2.9)}$$
\\
The quadratic fluctuations become 

$$ -R~ (T_1^*, T_2^*, T_3^*)  \left( \begin{array}{ccc} P^2 cos^2\theta&-(PQ-2\pi iz^2)cos^2\theta&0\\ -(QP+2\pi iz^2)cos^2\theta&Q^2&0\\0&0&P^2cos^2\theta+Q^2\end{array} \right)  \left( \begin{array}{c} T_1\\T_2\\T_3\end{array} \right), \eqno{(2.10)}$$
\\
in which $P = P_1-P_2$ and $Q = Q_1-Q_2$.   The  operator ``$Tr$'' was transformed to the  integration over variable $P_1$ and $P_2$ as that in [18].   Note that when $\theta =0$ then the above relation reduces to the $r=0$ limit of eq.(3.13) in [18].

   Now, after defining

 $$a = (Q + iP) / \sqrt{2(2\pi z^2)}, ~~~a^\dagger = (Q - iP) / \sqrt{2(2\pi) z^2}, \eqno{(2.11)}$$
which  satisfies $[a, a^\dagger]=1$, and defining

$$A= {1-cos\theta \over \sqrt {4 cos\theta}} a^\dagger + {1+cos\theta \over \sqrt {4 cos\theta}} a,  \eqno{(2.12)}$$
\\
which  also satisfies $[A, A^\dagger]=1$, then eq.(2.10) can be expressed as 

$$  -R ~ cos\theta~ 4\pi z^2 ~(\tilde T_1^*, \tilde T_2^*, \tilde T_3^*) \left( \begin{array}{ccc} (A^\dagger A-1)&A^\dagger A^\dagger &0\\ AA&(A^\dagger A+2)&0\\0&0&(2A^\dagger A+1) \end{array} \right)  \left( \begin{array}{c} \tilde T_1\\ \tilde T_2\\ \tilde T_3\end{array} \right),  \eqno{(2.13)}$$
\\
in which 

$$\left( \begin{array}{c} \tilde T_1\\ \tilde T_2\\ \tilde T_3\end{array} \right) = \left( \begin{array}{ccc}{1\over \sqrt{2}} & -{i\over \sqrt{2}} & 0
\\ -{1\over \sqrt{2}} & -{i\over \sqrt{2}} & 0 \\ 0 & 0 & 1 \end{array} \right)\left( \begin{array}{c}  T_1\\  T_2\\  T_3\end{array} \right).  \eqno{(2.14)}$$
\\
It is seen that without the $cos\theta$ term eq.(2.13) is just the $r=0$ limit of eq.(3.15) in [18] which investigates the  membrane-anti-membrane system.   This fact can be explained in the following.

   It is known [18] that for a pair of membranes the terms calculated using the Hamiltonian would depend only on $Q_1-Q_2$ and $P_1+P_2$, which will commute with each other. Thus the quadratic fluctuation for a pair of membranes vanishes.   For the intersecting branes system, we can divide the Matrix field as the anti-parallel part (which has a factor $cos\theta$) and the parallel part (which has a factor $sin\theta$).   Now as the parallel part (which behave as a pair of  branes) does not contribute to the quadratic fluctuation we are left only the anti-parallel part and thus a factor $cos\theta$ appears. 

   Notice that although eq.(2.13) has an overall factor $cos\theta$ the eq.(2.10) does not show such a factorizable property.

   To proceed we need to find the spectrum of mass operator, denote as $M_2$, of  (2.13).   This work has been done in [18].   But, for completeness, let us make a short review.   We can first define a basis of harmonic oscillator eigenfunctions $L_n$ satisfying $A L_n = \sqrt{n} L_{n-1}$ and $ A^\dagger L_n = \sqrt{n+1}L_{n+1}$. The eigenvectors are then of the form $(\alpha L_n, \beta L_{n-2}, \gamma L_{n-1})$.  Next, we choose $M_2$ to act on a vector of this form, then the eigenvector is transformed into a vector of the same type, with the matrix $\tilde{M_2}$ acting on $(\alpha,\beta,\gamma)$, where

$$\tilde M_2 = 4\pi R z^2 cos\theta~\left( \begin{array}{ccc}n-1 & \sqrt{n(n-1)} &0\\ \sqrt{n(n-1)} & n & 0\\ 0 &0 & 2n-1\end{array} \right).  \eqno{(2.15)}$$
\\
The above matrix as  one eigenvector $v_1 =(-\sqrt{n},\sqrt{n-1},0)$ with eigenvalue $0$, and two eigenvectors,  $v_2=(0,0,\sqrt{n})$ and $v_3=(\sqrt{n(n-1)},n,0)$, which have the same eigenvalue $4\pi z^2 R(2n-1)cos\theta$.   Note that the these eigenfunctions exist for any $n \ge 2$.   When  $n = 0$ eq.(2.15) implies that there is an eigenvector $v_0 =(1,0,0)$ with eigenvalue $``-4\pi z^2 Rcos\theta"$, which is a tachyon mode.    When  $n = 1$ eq.(2.15) implies that there is an eigenvector $v_4 =(1,0,0)$, with eigenvalue $0$ and another eigenvector $v_5 =(0,0,1)$ with eigenvalue $4\pi z^2 Rcos\theta$.  

   For the six other bosonic fluctuations $A_\mu$, $\mu=4,..., 9$, there will appear the terms

$$A_\mu^\dagger\left(P^2cos^2\theta+Q^2\right)  A_\mu = A_\mu^\dagger \left[cos\theta\left(2A^\dagger A+1\right)\right] A_\mu , \eqno{(2.16)}$$
\\
in which $A^\dagger$ and $A$ are defined in (2.12).   The eigenfunctions are  $L_n$ which have  eigenvalues  $(2n+1)cos\theta$ for any $n \ge 0$.   The fermion parts can be considered in the same way and there are four states with eigenvalue $(2n+2)cos\theta$ and four states with eigenvalue $2n~cos\theta$ for  $n \ge 0$.   These modes are irrelevant to the tachyon condensation.

\section {Branes Dynamics: Tachyon Condensation}
To consider the tachyon condensation let us first express (2.9) in terms of the new field $\tilde T_i$ defined in (2.14)
$$Tr [A_i, A_j] [A_i, A_j] = -4 \left[(\tilde T_1^\dagger \tilde T_1 + \tilde T_2^\dagger \tilde T_2 )^2 + 2 (\tilde T_1^\dagger \tilde T_3 - \tilde T_3^\dagger \tilde T_1)(\tilde T_2^\dagger \tilde T_3 - \tilde T_3^\dagger \tilde T_2)\right].\eqno{(3.1)}$$
As the tachyon mode found in the previous section has the eigenvector $v_0 =(1,0,0)$ and the eigenvalue $``-4\pi z^2 Rcos\theta``$, we see that the corresponding classical potential can be expressed as
$$V_{tachyon} = - 4\pi z^2 R~\tilde T_1^{(tachyon)\dagger} \tilde T_1^{(tachyon)} cos\theta + R~ (\tilde T_1^{(tachyon)\dagger} \tilde T_1^{(tachyon)})^2, \eqno{(3.2)}$$
in which the tachyon mode function is denoted as $\tilde T_1^{(tachyon)}$.  From the above potential we see that the tachyon will roll from the locally maximum point $\tilde T_1^{(tachyon)} =0 $ to the minimum point $\tilde T_1^{(mim)}$  with 

$$\tilde T_1^{(mim)} =  \sqrt{2\pi  z^2 cos\theta}.  \eqno{(3.3)}$$
\\
At the minimum point the tachyon field  become massless and stable.    The shifted field is denoted as new field $\tilde T_1^{(stable)}$ to which the other modes is coupled. 

   We now use the above result to explain how  the tachyon condensation appears in the matrix theory.

   First, from  (3.3) and (2.14) we have the relation

$$T_1^{(mim)} = {\sqrt 2\over 2} \sqrt{2\pi  z^2 cos\theta},\eqno{(3.4)}$$
$$T_2^{(mim)} = {i \sqrt 2\over 2} \sqrt{2\pi  z^2 cos\theta}.\eqno{(3.5)}$$
\\
Then, for the intersecting branes (denoted in (2.2)) we have a tachyon mode between them.   We can denote this state by the matrix

$$X_1 = \left( \begin{array}{cc}P_1 sin\theta&T_1^{(tachyon)}\\T_1^{(tachyon)\dagger}&P_2 sin\theta  \end{array} \right), ~~~ X_2 = \left( \begin{array}{cc}P_1 cos\theta&T_2^{(tachyon)}\\
T_2^{(tachyon)\dagger}& - P_2 cos\theta  \end{array} \right), \eqno{(3.6)}$$
\\
As the tachyon mode $T_i^{(tachyon)}$ in the above matrix will roll down to the minimum values  $T_i^{(min)}$ of (3.4) and (3.5), we see that eq.(3.6) can be expressed as

$$X_1 = \left( \begin{array}{cc}P_1 sin\theta&T_1^{(min)}\\T_1^{(min)\dagger}&P_2 sin\theta  \end{array} \right) + \left( \begin{array}{cc}0&T_1^{(stable)}\\T_1^{(stable)\dagger}&0  \end{array} \right),  \eqno{(3.7)}$$

$$X_2 = \left( \begin{array}{cc}P_1 cos\theta&T_2^{(min)}\\T_2^{(min)\dagger}&- P_2 cos\theta  \end{array} \right) + \left( \begin{array}{cc}0&T_2^{(stable)}\\T_2^{(stable)\dagger}&0  \end{array} \right),  \eqno{(3.8)}$$
\\
in which 
$$T_i^{(stable)} = T_i^{(tachyon)} - T_i^{(min)}.  \eqno{(3.9)}$$
As the second terms in (3.7) and (3.8) represent the interaction coming from the field $T_i^{(stable)}$ which are the stable states after the tachyons have rolled down to its stable point, the first terms therein therefore represent the tachyon condensation configurations of the intersecting branes.    To find the new condensed configuration we shall  diagonalize the first matrices in (3.7) and (3.8).   The physical interpretation of the diagonalization procedure is that we need to find a new eigenstate in which the diagonal elements are the associated eigenvalues.   In the same way, we shall now make a suitable linear combination of brane one and brane two (which are intersecting to each other) in order to find the new configurations which, as the interaction (i.e., the off-diagonal element in the matrix theory) has not yet been involved, will be independent and become diagonal in the matrix theory.  This interpretation is consistent with the fact that the new recombined brane has one half from an original brane and one half from the another original brane.

    We are now at the final step.  From eqs. (3.4), (3.5) , (3.7) and (3.8) we have to diagonalize the following two matrices  
$$\left( \begin{array}{cc}P_1 sin\theta&{\sqrt 2\over 2} \sqrt{2\pi  z^2 cos\theta}\\{\sqrt 2\over 2} \sqrt{2\pi  z^2 cos\theta}&P_2 sin\theta  \end{array} \right), \eqno{(3.10)}$$

$$\left( \begin{array}{cc}P_1 cos\theta&{i \sqrt 2\over 2} \sqrt{2\pi  z^2 cos\theta}\\{-i\sqrt 2\over 2} \sqrt{2\pi  z^2 cos\theta}& -P_2 cos\theta  \end{array} \right). \eqno{(3.11)}$$
\\
The associated eigenvalues $x_d$ and $y_d$ are 
$$x_d = x_0 sin\theta \mp \sqrt{\pi z^2 cos\theta}.\eqno{(3.12)}$$
$$y_d = \mp \sqrt{x_0^2 cos\theta^2 + \pi z^2 cos\theta},\eqno{(3.13)}$$
in which we let $P_1=P_2=x_0$, which represents one of the coordinates of the branes system.  (The other coordinate is $Q_1= Q_2= y_0$ which is irrelevant to the system as the branes are translation invariant in this direction.)

   It is easy to see that the above two equations imply a simple relation

$$\left(x_d \mp \sqrt{\pi z^2 cos\theta}\right)^2 ={sin\theta^2\over cos\theta^2} \left(y_d^2 - \pi z^2 cos\theta\right),\eqno{(3.14)}$$
\\
and thus the recombined branes become an ``asymmetric'' hyperbola which is different from the symmetric one in the previous investigation [7-9].  The geometry of the recombination is shown in figure 1.  The reason why the recombined branes become ``asymmetric'' is not clear.   It seems that the gauge fields on the noncommutative branes will cause an extra effect of making the dash-line curve in figure 1 to be shifted to the solid-line curve.  It  remains to  be clarified in futher investigations.
\\

\hspace{4cm}\scalebox{0.7}{\includegraphics{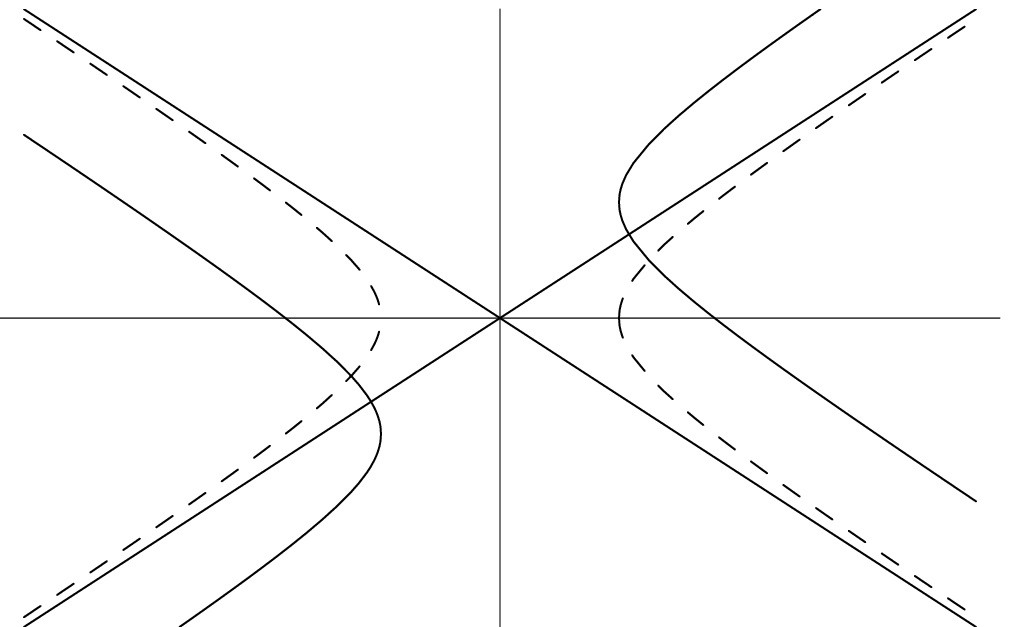}} 
\\
\\
{\it ~~~Fig.1. Recombination of intersecting noncommutative branes.   The original pair of branes intersect at (0,0).  After the recombination the new branes become the solid-line curves.  The dash-line curves represent those in the case of commutative branes.}

\section {Conclusion}

   The previous investigations of the mechanism of recombination of intersecting branes have used the super Yang-Mills theory [7,8] or the effective tachyon field theory [9], which have taken the truncated  Lagrangian and can, at most, regarded as a small angle approximation.   The another approach to the problem by using the world-volume field theory [12] can avoid the approximation.  M(artrix) theory as a fundamental theory [12,13] provide us with an interesting approach to investigate the tachyon condensation and recombination of intersecting  membranes.   It can also be used to describe the intersecting D-brane in the presence of non-zero gauge field, i.e., the non-commutative branes system.

   In this paper we have investigated the mechanism of recombination of intersecting noncommutative branes (or membranes) in the M(atrix) theory approach.   We find the spectrum of the off-diagonal fluctuations.   It is seen that the spectrum is like the membrane-anti-membrane system [18] while with an overall factor $cos\theta$.   The tachyon mode therein signals the instability of the intersecting branes.  We have detailed how to find the tachyon condensation configurations of the intersecting noncommutative D-branes in the M(atrix) theory.   Our result shows that the recombined brane configurations are different from those of the commutative cases studied in the previous literatures [7-9].   We argue that the gauge fields on the noncommutative branes will cause an extra effect to make such a difference.
Finally, it seems that our prescription can be extended to discuss other unstable noncommutative brane configurations, such as the brane ending on branes, brane system with different dimensions, etc.   We will investigate these problems in the future.

\newpage
\begin{enumerate}
\item A. Sen,  ``Tachyon Condensation on the Brane Antibrane System'', 
  JHEP 9808  (1998)  012,   hep-th/9805170;   ``Descent Relations Among Bosonic D-branes'',  Int.\ J.\ Mod.\ Phys.  A14 (1999) 4061,  hep-th/9902105;   ``Universality of the Tachyon Potential'',  JHEP  9912 (1999) 027, hep-th/9911116.
\item  M.~Berkooz, M.~R.~Douglas and R.~G.~Leigh, ``Branes Intersecting at Angles'', Nucl. Phys. B480 (1996) 265, hep-th/9606139; M. M. Sheikh Jabbari,``Classification of Different Branes at Angles'',  Phys. Lett. B420 (1998) 279, hep-th/9710121; N. Ohta and P. Townsend,``Supersymmetry of M-Branes at Angles'', Phys. Lett. B418 (1998) 77, hep-th/9710129.
\item D. J. Smith, ``Intersecting Brane Solution in string and M-Theory'', Class. Quant. Grav. 20 (2003) R233,  hep-th/0210157.
\item R.~Blumenhagen, B.~K\"ors, D.~L\"ust and T.~Ott,  ``The Standard Model from Stable Intersecting Brane World Orbifolds'', Nucl. Phys. B616 (2001) 3, hep-th/0107138; D.~Cremades, L.~E.~Ibanez and F.~Marchesano,
``Intersecting Brane Models of Particle Physics and the Higgs Mechanism'', 
JHEP 0207 (2002) 022, hep-th/0203160; C.~Kokorelis, ``Exact Standard model Structures from Intersecting D5-Branes'', hep-th/0207234. 
\item F. Marchesano, ``Intersecting D-brane Model'', hep-th/0307252.
\item J.~Garcia-Bellido, R.~Rabadan and F.~Zamora, ``Inflationary Scenarios from Branes at Angles'', JHEP 0201 (2002) 036, hep-th/0112147; R.~Blumenhagen, B.~K\"ors, D.~L\"ust and T.~Ott, ``Hybrid Inflation in Intersecting Brane Worlds'', Nucl. Phys. B641 (2002) 235, hep-th/0202124;
N.~Jones, H.~Stoica and S.~-H.~Henry Tye, ``Brane Interaction as the Origin of Inflation'',  JHEP 0207 (2002) 051,  hep-th/0203163; 
\item K.~Hashimoto and S.~Nagaoka, ``Recombination of Intersecting D-branes by Local Tachyon Condensation'', JHEP 0306 (2003) 034, hep-th/0303204; K.~Hashimoto and W. Taylor, ``String between Branes'',  hep-th/0307297. 
\item T. Sato, ``D-brane Dynamics and Creations of Open and Closed Strings after Recommbination'',  hep-th/0304237. 
\item Wung-Hong Huang, ``Recombination of Intersecting D-branes in  Tachyon Field Theory'', Phys. Lett. B564 (2003) 155, hep-th/0304171.
\item  B.\ Zwiebach,  ``A Solvable Toy Model for Tachyon Condensation in String Field  Theory'' ,  JHEP  0009  (2000)  028,  hep-th/0008227 ; J.\ A.\ Minahan and B.\ Zwiebach,  ``Field Theory Models for Tachyon and Gauge Field String  Dynamics'',  JHEP  0009 (2000) 029,  hep-th/0008231; Wung-Hong Huang, ``Brane-anti-brane Interaction under Tachyon Condensation.  Phys. Lett. B561 (2003) 153,  hep-th/0211127.
\item J. Erdmenger, Z. Gralnik, R. Helling, and I. Kirsch,  ``A World-Volum Perspective on the Recombination of Intersecting Branes'', hep-th/0309043.
\item T.~Banks, W.~Fischler, S.~Shenker and L.~Susskind, ``M-Theory as a
Matrix Model: A Conjecture'', Phys. Rev. D55 (1997) 5112, hep-th/9610043;
N.~Ishibashi, H.~Kawai, I.~Kitazawa, and A.~Tsuchiya, ``A large-N reduced model as superstring'', Nucl. Phys. B492 (1997) 467; hep-th/9612115. 
\item W. Taylor, ``M(artrix) Theory: Matrix Quantum Mechanics as a Fundamental Theory'',  Rev. Mod. Phys. 73 (2001) 419, hep-th/0101126.
\item P. K. Townsend,  ``D-branes from M-branes'', Phys. Lett. B373 (1996) 68, hep-th/9512062; T. Banks, N. Seiberg and  S. Shenker,  ``Branes from Matrices'',  Nucl. Phys. B490 (1997) 91; hep-th/9612157; E.~Keski-Vakkuri and P.~Kraus, ``Notes on Branes in Matrix Theory," Nucl. Phys. B510 (1998) 199, hep-th/9706196.
\item M. R. Douglas; ``Branes within Branes'' in Cargese 97,  hep-th/9512077.
\item  N.~Seiberg and E.~Witten, ``String Theory and Noncommutative Geometry,'' , JHEP  9909 (1999) 032, hep-th/9908142.
\item K. Kaviani, S. Parvizi  and A. H. Fatollahi, ``Interaction of Branes at Angles in M(atrix) Theory'', Phys. Lett. B439 (2000) 271, hep-th/9808046.
\item  O.~Aharony and M.~Berkooz, ``Membrane Dynamics in M(atrix) Theory," Nucl. Phys. B491 (1997) 184, hep-th/9611215.
\item   H.  Awata, S. Hirano and Y. Hyakutake``Tachyon Condensation and Graviton Production in Matrix Theory'', hep-th/9902158.
\item  M. M. Sheikh Jabbari,``Different D-brane interactions'',  Phys. Lett. B394 (1997) 288. 

\end{enumerate}

\end{document}